\documentstyle[twocolumn,aps,epsfig]{revtex}

\begin{document}

\draft

\title{Orthorhombic versus monoclinic symmetry of the 
charge-ordered state of NaV$_2$O$_5$}

\author{Sander van Smaalen$^*$, Peter Daniels and Lukas Palatinus}
\address{Laboratory for Crystallography, University of Bayreuth, 
D-95440 Bayreuth, Germany}

\author{Reinhard K. Kremer}
\address{Max Planck Institute for Solid State Research, 
Heisenbergstrasse 1, D-70569 Stuttgart, Germany}

\date{\today}
\maketitle

E-mail: smash@uni-bayreuth.de 

\begin{abstract}
High-resolution X-ray diffraction data show that the 
low-temperature superstructure of $\alpha'$-NaV$_2$O$_5$ 
has an $F-$centered orthorhombic $2a \times 2b \times 4c$ 
superlattice. 
A structure model is proposed, that is characterized by 
layers with zigzag charge order on all ladders 
and stacking disorder, 
such that the averaged structure has space group Fmm2. 
This model is in accordance with both X-ray scattering 
and NMR data. 
Variations in the stacking order and disorder offer an 
explanation for the recently observed devils staircase  
of the superlattice period along $c$. 
\end{abstract}


The low dimensional transition metal oxide $\alpha'$-NaV$_2$O$_5$ 
undergoes a phase transition at a temperature of T$_c = 34$ K. 
The transition is characterized by the development of both 
a nonmagnetic groundstate and a superstructure \cite{isobem1996,fujii1997}. 
General agreement exists that the phase transition is 
associated with the development of charge order on the 
vanadium sublattice \cite{mostovoy1998,seoh1998}, but 
the mechanism of the transition has not been revealed yet. 

At room temperature $\alpha'$-NaV$_2$O$_5$ crystallizes 
in space group Pmmn 
\cite{vonschnering1998,meetsma1998,smolinski1998}. 
There is one crystallographically independent vanadium 
atom, that is in the mixed-valence state $4.5+$. 
The structure can be considered as built of layers of 
two-leg ladders V$_2$O$_3$, that are stacked along $\vec{c}$, 
alternating with sodium atoms and additional oxygen. 
The lattice parameters of the basic structure at 15 K are 
$a=11.294$ \AA{}, $b = 3.604$ \AA{}, 
and $c = 4.755$ \AA{} \cite{ludecke1999a}. 
The superlattice below T$_c$ can be described by an 
$F-$centered orthorhombic $2a \times 2b \times 4c$ 
supercell. 
The superstructure was found to have symmetry $Fmm2$, 
but it showed two peculiar features 
\cite{ludecke1999a,vansmaalen2000a}: 
(i) In one layer ladders with zigzag charge order 
alternate with ladders with vanadium in the mixed-valence 
state, 
(ii) Each of the two consecutive layers contains half 
of the six crystallographically independent vanadium 
atoms, but their structures were nearly equal. 
This crystal structure was found to be in agreement 
with two other X-ray diffraction measurements 
\cite{deboer2000,chatterji2001}. 

Theoretical analyses have produced models that show 
zigzag charge order on all ladders 
\cite{mostovoy1998,seoh1998,riera1999,gros1999,yaresko2000,thalmeier2000,langari2001}. 
However, most approaches did not consider the true 
supercell, and therefore they cannot be expected to 
reveal all aspects of the mechanism of the phase transition. 

Various experiments, including 
anomalous X-ray scattering \cite{nakao2000}, 
inelastic neutron scattering \cite{grenier2001}, 
Raman spectroscopy \cite{konstantinovic1999}, 
and NMR \cite{fagot-revurat2000,ohama2000a}, 
have suggested zigzag charge order on all ladders. 
Such a model is at variance with the published 
crystal structure, and it is not possible for 
any ordered structure with Fmm2 symmetry \cite{ludecke1999a}. 
Most notably, $^{23}$Na NMR has found eight resonances, 
that were interpreted as being due to eight 
crystallographically independent atomic sites, 
whereas the crystal structure in Fmm2 only has six 
indepent Na sites. 
It was proposed that the true symmetry 
of the low-temperature structure might be a 
subgroup of Fmm2 corresponding to the loss of the 
F-center \cite{ohama2000a}. 
Alternatively, monoclinic symmetry was considered 
\cite{konstantinovic1999}. 

In order to determine the true superstructure of 
NaV$_2$O$_5$, we have measured high-resolution, 
high-sensitive synchrotron radiation X-ray diffraction. 
The experiment indicates that the true superlattice 
is $F-$centered on the $2a\times 2b\times 4c$ supercell. 
We show that an all zigzag charge order model 
with orthorhombic symmetry is possible assuming 
stacking disorder. 
This model is in agreement with both X-ray diffraction
and NMR. 

X-ray diffraction experiments were performed at beamline 
ID10A of the ESRF in Grenoble, France. 
Monochromatic radiation of a wavelength of 
$\lambda = 0.66057$ \AA{} was selected by the 220 reflection 
of diamond. 
Bragg reflections were measured by $\omega-$scans 
using a scintillation detector. 

NaV$_2$O$_5$ single crystals were grown by the flux method 
(batch number $E106$ \cite{johnstondc2000}). 
A crystal of dimensions 
$0.05 \times 0.06 \times 0.13$ mm$^3$ was mounted 
on a closed-cycle cryostat placed on a Huber diffractometer. 
The temperature was checked by measuring the intensity of 
a strong satellite reflection, that was found to be present 
for T $< 33 \pm 1$ K only. 
The strong Bragg reflections could be indexed on the basis 
of the small primitive orthorhombic unit cell, in agreement 
with the literature. 

\begin{figure}
\noindent 
\epsfig{file=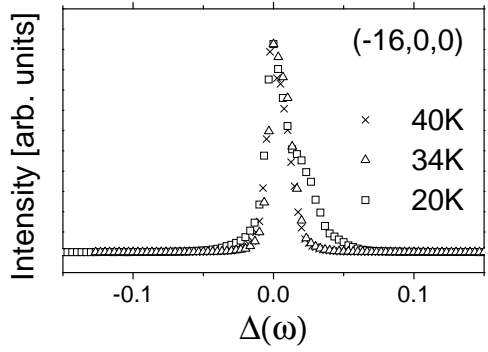,width=7.0cm,angle=-90}\\{} 
\epsfig{file=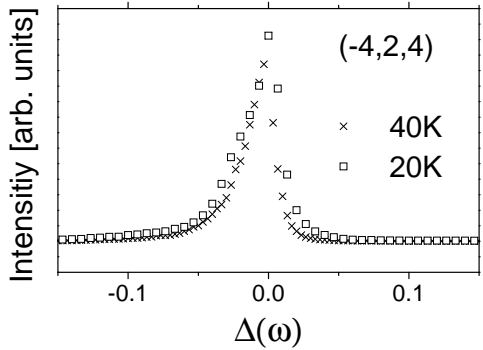,width=7.0cm,angle=-90}\\{}
\epsfig{file=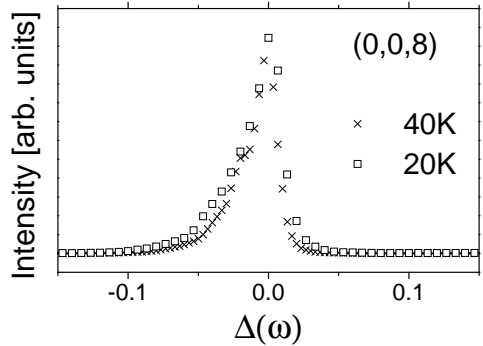,width=7.0cm,angle=-90}
\caption{Intensity against crystal orientation ($\omega-$scan) 
for selected main reflections measured below and 
above the phase transition. 
(a) The $(-16,0,0)$ reflection; 
(b) $(-4,2,4)$; 
(c) $(0,0,8)$. 
Note that reflection indices refer to the supercell.}
\label{f-profiles}
\end{figure}

A possible monoclinic distortion of the superlattice 
would result in a domain structure, that gives rise 
to split Bragg reflections, with the splitting angle 
equal to twice the deviation of the monoclinic angle from 
90 deg. 
In order to test the hypothesis of a monoclinic lattice, 
a series of main reflections was measured at temperatures 
of both 20 K and 40 K. 
Because of limitations of the cryostat not all directions 
could be reached, but the measured reflections test any 
possible monoclinic distortion with the c axis as unique 
axis as well as most other possible lattice distortions. 
The profiles were found to broaden slightly below the 
phase transition, but splitting was not observed (Fig. \ref{f-profiles}). 
This limits a possible 
lattice distortion (e.g. monoclinic angle) to the 
Half Width at Half Maximum (HWHM) of the reflections, 
i.e. to $0.009$ deg. 
Furthermore, the changes of profiles were also found 
for the $(0,0,l)$ reflections (Fig. \ref{f-profiles}c), 
that should have remained sharp for a 
monoclinic distortion with the $c-$axis as unique direction. 
Therefore, we conclude that the lattice is not monoclinic. 

In a second experiment reflections corresponding to a 
primitive $2a\times 2b\times 4c$ supercell were measured 
at 20 K. 
Significant intensities were found for all eight measured 
first-order satellites as well as for all eight measured 
second-order satellites. 
Except for the forbidden (11,0,3) reflection, 
scattered intensity was not found for all measured $51$ reflection 
positions that were forbidden by the $F-$center. 
However, the intensity of this forbidden reflection 
was the same at 20 K and 40 K, 
and its presence is not related to the phase transition. 
Presently we have achieved a much higher sensitivity 
towards weak scattering effects than in our previous 
experiment \cite{ludecke1999a}.
It is characterized by the ratio of the intensity of 
$25000$ counts/s in the maximum of the $(-21,3,5)$ first-order 
satellite and the intensity of $6$ counts/s in the background. 

In view of these results, we have re-analyzed the 
low-temperature structure assuming various symmetries 
of the $F-$centered $2a\times 2b\times 4c$ supercell. 
The data by Bernert {\it et al.} \cite{chatterji2001} appear 
to form the most complete set,
and they have been used for all refinements presented here. 
In addition to data averaged in $mmm$ Laue symmetry 
(denoted as orthorhombic data), we have used the 
same intensities averaged in point group $\bar{1}$ 
(triclinic data). 

Refinement of the orthorhombic superstructure (space group 
$Fmm2$) against orthorhombic data reproduced the model 
by L\"udecke {\it et al.} \cite{ludecke1999a}. 
The same structure is obtained from the refinement of 
the orthorhombic model against the triclinic data, 
although the $R$ value now is higher (Table \ref{t-r-factors}). 
Assuming twinning, refinements with structure models according to 
$F112/d$ or $F11d$ (standard settings $A2/a$ and $Aa$) 
gave $R$ factors that were higher than for the $Fmm2$ 
structure. 

\begin{figure}
\noindent 
\epsfig{file=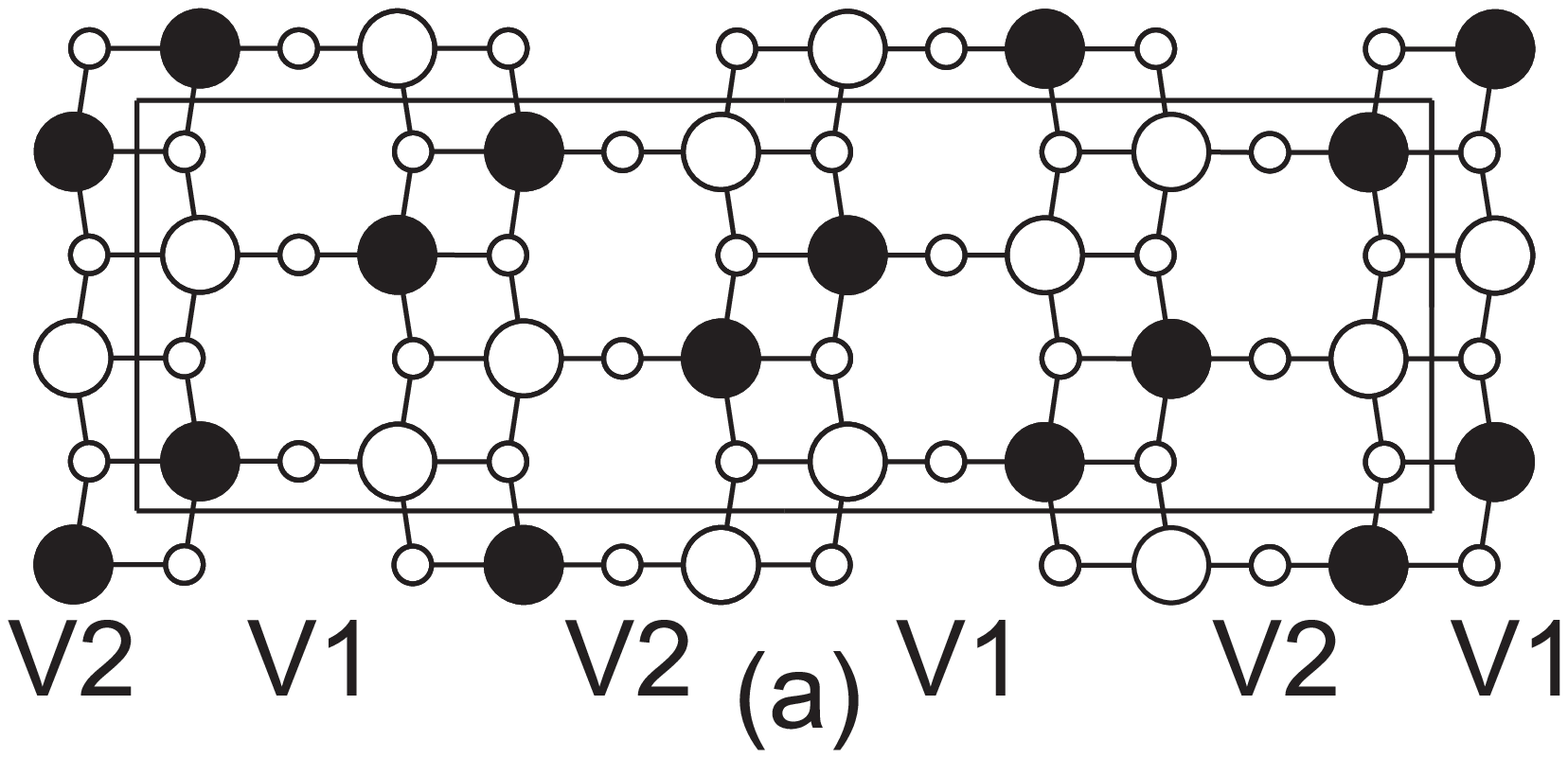,width=7.0cm} \\{} 
\epsfig{file=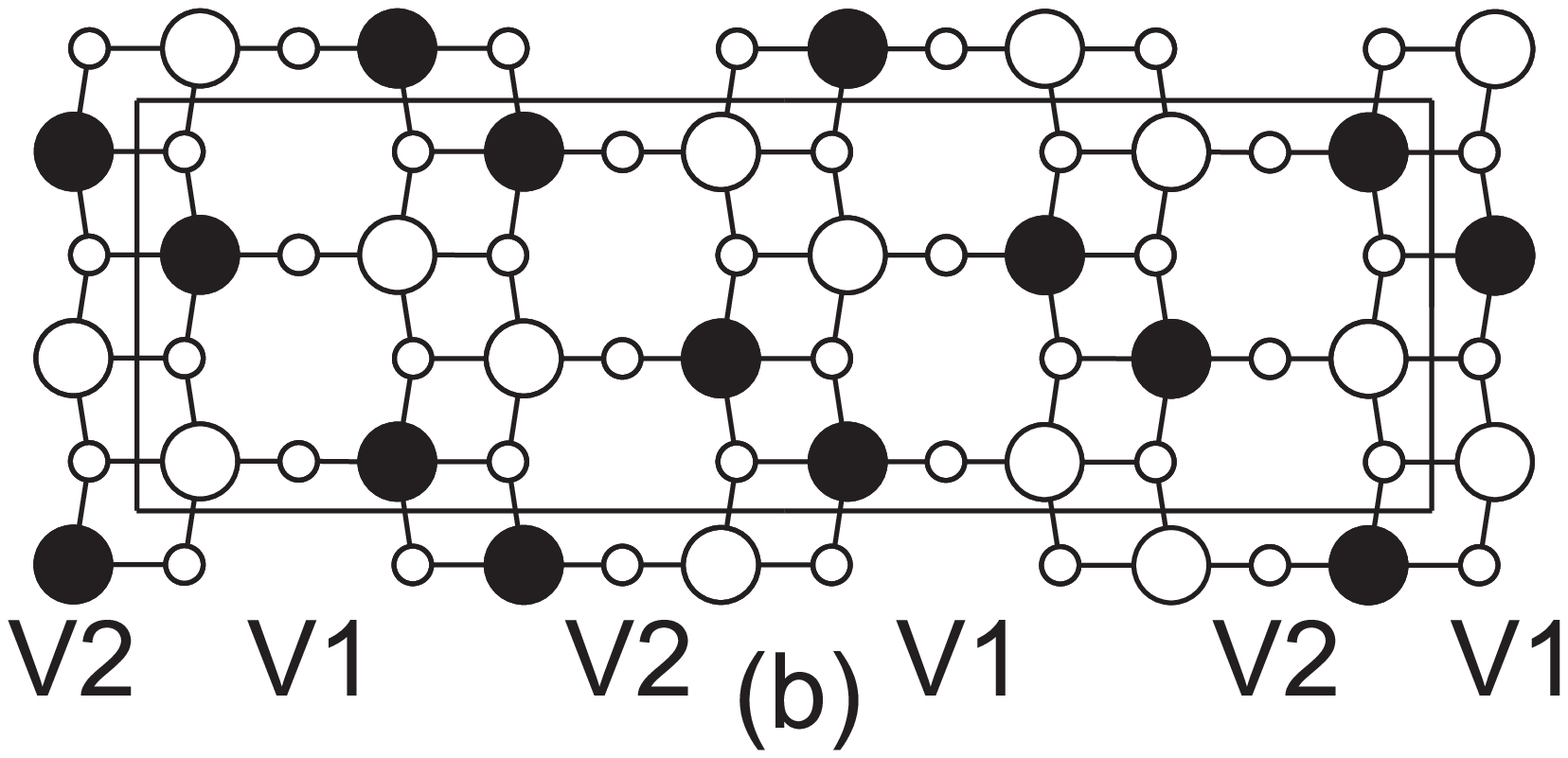,width=7.0cm}  
\caption{\label{f-super_layers}
The projection of one layer of the 
superstructure of NaV$_2$O$_5$ with charge order 
according to [3]. 
The $2a\times 2b$ supercell is indicated. 
Large filled and open circles represent vanadium atoms 
in the $4+\delta$ and $5-\delta$ valence states, 
respectively [9]. 
Small circles represent oxygen atoms. 
(a) Position {\bf A} of the charge order. 
Position {\bf B} is obtained from {\bf A} by a shift 
over b (half the superlattice constant). 
(b) Position {\bf D} of the charge order, 
that is related to {\bf A} by a shift over 
b of the V1 type ladders only. 
Position {\bf C} is related to {\bf A} by a shift over 
b of the V2 type ladders only. }
\end{figure}

Refinement of the monoclinic structure 
with space group $F112$ (standard setting $A2$) against 
triclinic data leads to $R$ = 0.074 
and a volume ratio of the twins equal to $0.75$. 
The modulation of the V2 type of atoms is slightly 
smaller than in the $Fmm2$ model, while the V1 type of 
atoms have shifts of less than one third of the shifts 
of the V2 type atoms (Fig. \ref{f-super_layers}) \cite{ludecke1999a}. 
This model does not meet the requirements of 
similar zigzag charge order on all ladders. 
Most likely, the shifts of the V1 type atoms represent 
a fit to errors in the data. 
A structure with all ladders equal is obtained by 
setting the shifts of the V1 type atoms equal to those 
of the V2 type atoms. 
The refinement now converges at 
$R$ =0.082 and a twin volume ratio of $0.98$. 
Almost perfect correlations are found between 
the parameters. 
It thus appears that an infinite number of monoclinic 
structure models give the same fit to the data as the 
orthorhombic structure $Fmm2$, including a monoclinic 
structure $F112$ with equal zigzag charge order on all ladders. 

Assuming four twin domains, refinements in triclinic $F\bar{1}$ 
symmetry converged at  $R$ = 0.075, with zigzag 
charge order on all ladders. 
Thus a better fit to the data was obtained than in 
the orthorhombic structure model. 
There are eight crystallographically independent 
vanadium atoms, but there are only four independent 
sodium atoms. 
Despite the good fit to the diffraction data, this 
model is not in agreement with the observations made 
by NMR \cite{fagot-revurat2000,ohama2000a}. 

The only possibility for complete zigzag charge order 
within the orthorhombic symmetry is disorder. 
For this, we consider the superstructure of a single 
layer as given by Mostovoy and Khomsky \cite{mostovoy1998}. 
Given the $2a\times 2b$ supercell, there are four 
equivalent realisations of this superstructure, that 
we denote by {\bf A}, {\bf B}, {\bf C}, and {\bf D} in 
a manner similar to the notation for different stacking 
sequences of a close packed structure of spheres 
(Fig. \ref{f-super_layers}). 
If we superimpose a layer {\bf A} with either {\bf C} 
or {\bf D}, an averaged structure results in which 
every other ladder is non-modulated. 
This averaged structure precisely is the 
structure of a single layer within the refined 
$Fmm2$ superstructure model \cite{ludecke1999a}. 
Thus a disordered stacking of Layers {\bf A} and {\bf C} 
(or equivalently {\bf A} and {\bf D}, {\bf B} and {\bf C} 
or {\bf B} and {\bf D}) results in 
a structure with an average unit cell 
$2a\times 2b\times c$ in which the structure of 
the single layer corresponds to the observed structure 
of the individual layers. 
Of course, this model of stacking disorder is too 
simple, as it does not explain the observed 
quadrupling of the $c$ lattice parameter.  

Stacking faults can be considered within an ordered 
superstructure with orthorhombic symmetry Ccc2 on the 
$2a\times 2b\times 4c$ supercell. 
Possible stacking sequences in Ccc2 are {\bf ADCB} and {\bf ABCD}, 
whereby we have imposed the condition that neighboring 
layers must be different. 
A sequence with one stacking fault can be 
\begin{displaymath} 
\cdots{\bf ADCB\: ADCB ^{\bullet} DABC\: DABC}\cdots
\end{displaymath} 
where a dot $^{\bullet}$ denotes the position of the 
stacking fault. 
On the average this structure has a 
$2a\times 2b\times 4c$ supercell with stacking sequence 
\begin{displaymath} 
<{\bf A,D}> <{\bf D,A}> <{\bf C,B}> <{\bf B,C}> 
\end{displaymath} 
where $<{\bf A,D}>$ denotes one layer with a structure that 
is the average of the structures of the layers {\bf A} 
and {\bf D}, and $<{\bf A,D}>$ = $<{\bf D,A}>$. 
This averaged structure precisely is the structure with 
space group $Fmm2$ as previously reported in \cite{ludecke1999a}. 
Because the Bragg reflections in X-ray scattering reflect 
only the averaged structure, a model of layers with 
zigzag charge order on all ladders, but with 
the appropriate stacking disorder is in complete 
accordance with the measured diffraction intensities. 
Refinements with shifts of the V1 type atoms 
according to this disorder model indeed gave the same 
$R$ values as the ordered $Fmm2$ model.
The lattice is orthorhombic, and the 
disorder model is in accordance with our failure to 
observe any splitting of Bragg reflections. 
It is noticed, that the stacking disorder given above 
is just one example of how the observed average structure 
can be obtained. 
The true modes of stacking disorder should follow from 
the analysis of diffuse scattering or theoretical 
modelling. 

The proposed stacking disorder of layers with full 
zigzag charge order is in agreement with all available 
experimental data. 
It explains both X-ray diffraction data and NMR. 
The reasons for the stacking disorder will lie in 
the multiple minima of the superstructure, and the 
resulting frustration. 
Considering nearest neighbor contacts only, 
the layer structures {\bf A}, {\bf B}, {\bf C}, 
and {\bf D} are equally probable, and stacking 
sequences {\bf AD}, {\bf AC}, {\bf BD}, and 
{\bf BC}  have the same energy. 
The notion of different stacking sequences with nearly 
equal energies offers an explanation for the recently 
observed variation of the superlattice length 
along $\vec{c}$ \cite{ohwada2001}. 
The different superstructures observed when applying 
hydrostatic pressure are to be considered 
as the result of different stacking sequences. 

In conclusion, we have found that the true global 
symmetry of the low-temperature superstructure of 
NaV$_2$O$_5$ is Fmm2 on a $2a\times 2b\times 4c$ supercell. 
From the X-ray data, there is no direct evidence for another 
structure model than the fully ordered superstructure with 
alternatingly charge ordered and mixed valence ladders 
as given in Ref. \cite{ludecke1999a}. 
A monoclinic distortion is ruled out, 
while the ordered structure with triclinic symmetry 
did not explain the NMR data of Refs. \cite{fagot-revurat2000,ohama2000a}. 
In order to accommodate observations by experimental 
techniques other than X-ray scattering, we propose that 
the true superstructure might be composed of layers 
with zigzag charge order on all ladders, that shows 
stacking disorder within a superstructure of orthorhombic 
symmetry, e.g. within a model with the space group $Ccc2$. 
This model explains all presently available experimental 
information. 
Furthermore, it provides an explanation for the 
observation of the devils staircase behavior of 
the superlattice parameter along $\vec{c}$ under 
pressure. 
Finally it is noted, that the presence of stacking 
disorder in the superstructure might be the origin 
of the non-standard value of the critical exponent 
of the order parameter \cite{ravy1999,gaulin2000}, 
and of the splittings observed in the anomaly of 
the heat capacity at the phase transition \cite{johnstondc2000}. 

X-ray scattering experiments were performed at beamline 
ID10A of the European Synchrotron Radiation Facility 
(ESRF) in Grenoble, France (Experiment Number HS-1427). 
We greatfully acknowledge the assistance of the 
beamline staff, and in particular of Dr. F. Zontone. 
We thank E. Br\"ucher for help with the crystal growth, 
and T. Chatterji for making his scattering data available. 
Financial support was obtained from the German Science Foundation 
(DFG) and the Fonds der Chemischen Industrie (FCI). 


 
%

\begin{table}
\caption{Partial reliability factors (R-factors) between 
observed and calculated superlattice reflections for 
various structure models and two data sets. 
Lower values indicate better agreements. 
}
\label{t-r-factors}
\begin{tabular}{l l l l l}
Structure model & \parbox{2.0cm}{orthorhombic data} & 
\parbox{2.0cm}{triclinic data} \\
\hline 
$Fmm2$             & 0.063 & 0.082 \\ 
$F112/d$ (twinned) & 0.138 & 0.145 \\
$F11d$ (twinned)   & 0.121 & 0.129 \\ 
\end{tabular}
\end{table}

\clearpage

\end{document}